%% file: Main.tex
\documentclass[sigconf]{acmart}
\usepackage{xcolor}
\usepackage{multirow}
\renewcommand{\arraystretch}{1.5}  

\newcommand{\edit}[1]{\textcolor{black}{#1}}
\newcommand{\note}[1]{\textcolor{black}{#1}}
\newcommand{\rev}[1]{\textcolor{black}{#1}}
\newcommand{\update}[1]{\textcolor{black}{#1}}

\AtBeginDocument{%
  }

\copyrightyear{2026}
\acmYear{2026}
\setcopyright{cc}
\setcctype{by}
\acmConference[CHIWORK '26]{Proceedings of the 5th Annual Symposium on Human-Computer Interaction for Work}{June 22--25, 2026}{Linz, Austria}
\acmBooktitle{Proceedings of the 5th Annual Symposium on Human-Computer Interaction for Work (CHIWORK '26), June 22--25, 2026, Linz, Austria}
\acmDOI{10.1145/3808045.3808050}
\acmISBN{979-8-4007-2429-9/2026/06}

\begin{document}

\title[The Social Dynamics of Developing Generative AI Literacy in the Workplace]{"If You're Very Clever, No One Knows You've Used It": The Social Dynamics of Developing Generative AI Literacy in the Workplace}

\author{Qing (Nancy) Xia}
\email{nancy.xia.18@ucl.ac.uk}
\orcid{}
\affiliation{%
  \institution{University College London}
  \city{London}
  \country{UK}
}

\author{Marios Constantinides}
\affiliation{%
  \institution{CYENS Centre of Excellence}
  \city{Nicosia}
  \country{Cyprus}}
  \authornote{Also affiliated with University of Cyprus, Cyprus and University College London, UK.}
\email{marios.constantinides@cyens.org.cy}

\author{Advait Sarkar}
\affiliation{%
  \institution{Microsoft Research}
  \city{Cambridge}
  \country{UK}}
  \authornote{Also affiliated with University of Cambridge, University College London, UK}
\email{advait@microsoft.com}

\author{Duncan P. Brumby}
\affiliation{%
 \institution{University College London}
 \city{London}
 \country{UK}}
\email{d.brumby@ucl.ac.uk}

\author{Anna L. Cox}
\affiliation{%
  \institution{University College London}
  \city{London}
  \country{UK}}
  \authornote{Also affiliated with University of Bristol, UK and CYENS Centre of Excellence, Cyprus}
\email{anna.cox@ucl.ac.uk}

\renewcommand{\shortauthors}{Xia et al.}

\begin{abstract}

Generative AI (GenAI) tools are rapidly transforming knowledge work, making AI literacy a critical priority for organizations. However, research on AI literacy lacks empirical insight into how knowledge workers’ beliefs around GenAI literacy are shaped by the social dynamics of the workplace, and how workers learn to apply GenAI tools in these environments. To address this gap, we conducted in-depth interviews with 19 knowledge workers across multiple sectors to examine how they develop GenAI competencies in real-world professional contexts. We found that, while knowledge sharing from colleagues supported learning, the ability to remove cues indicating GenAI use was perceived as validation of domain expertise. These behaviours ultimately reduced opportunities for learning via knowledge sharing and undermined transparency. To advance workplace AI literacy, we argue for fostering open dialogue, increasing visibility of user-generated knowledge, and greater emphasis on the benefits of collaborative learning for navigating rapid technological developments.


\end{abstract}




\maketitle

\input{01_Introduction}

\input{02a_GenAI_in_Knowledge_Work}
\input{02b_Digital_skills_in_workplace}

\input{02c1_GenAI_Literacy}

\input{03_Method}

\input{04_2_Results}

\input{05_Discussion}

\begin{acks}
We are grateful for the funding provided by the Engineering and Physical Sciences Research Council [Grant number: EP/W522077/1] and Microsoft Research through its EMEA PhD Scholarship Programme, and the participants for their time and participation. We also thank the anonymous reviewers for their constructive and insightful feedback which helped improve the clarity and presentation of the paper. 
\end{acks}

\balance
\bibliographystyle{ACM-Reference-Format}
\bibliography{AI_Literacy_Ref}

\appendix

\end{document}

%% file: 01_Introduction.tex
\section{Introduction}

Generative Artificial Intelligence (GenAI) tools are rapidly redefining modern knowledge work  \cite{eloundou2023, felten2023, chui2023, hatzius2023}. Adoption rates have increased annually since 2023 \cite{singla2024, singla2025}. Around 78\% of 31,000 knowledge workers surveyed in Microsoft and LinkedIn's 2024 Annual Report of Work Trends Index reported bringing their own AI tools to work \cite{microsoft2024}. As a result, there has been increased interest in reskilling existing employees to newly automated workflows, and ensuring responsible usage by increasing users' awareness of the social and ethical implications of GenAI use \cite{singla2025, singla2024, nguyen2025, mayer2025, zirar2023}.

Establishing AI literacy is now essential. AI literacy describes the competencies necessary for individuals to understand, interact with, and critically evaluate AI technologies \cite{long2020, pinski2024, almatrafi2024}. Important competencies often include understanding the technical underpinnings of AI \cite{long2020, carolus2023}; interaction techniques (e.g., prompt design \cite{korzynski2023}); variations of learning, critical thinking, reflection, and metacognitive skills \cite{lee2024a, ng2021, pinski2024, markauskaite2022, cox2024}; and socio-technical competencies, such as understanding the role of human actors in AI use through the lens of social and ethical impacts (e.g., \cite{carolus2023, annapureddy2025, pinski2024}). Lower levels of AI literacy are consistently linked with a range of problems, including a tendency to over-rely on AI tools \cite{zhan2024, tully2025, popa2024}, be more susceptible to hallucinations and biases \cite{peng2023}, lower job performance compared to peers with higher AI literacy \cite{liu2025a}, and lower employability \cite{portocarreroramos2025}. 

Significant effort has been directed towards defining and designing for student AI literacy in formal educational environments \cite{ng2021, chiu2024, yang2025}. However, professional users often rely on more unstructured methods of learning (e.g., informal knowledge sharing \cite{kiani2020, sarkar2018, lemmetty2020a}) to develop highly personalised understandings of how and when different tools should be applied in their work, that can differ from those emphasised in existing AI literacy frameworks. Indeed, current GenAI adoption among employees often occurs independently of organisational guidance and strategy: surveys indicate that 53\% of employees do not disclose their GenAI use to their employers \cite{microsoft2024}, while McKinsey \& Company's 2025 analysis showed that C-suite leaders significantly underestimated the extent of such adoption in day-to-day work \cite{mayer2025}. Many professionals are therefore constructing their own understanding of GenAI competencies and learning strategies, shaped less by organisational guidance or formal literacy frameworks, and more by personal priorities and their wider social context \cite{heyder2021}. 

As the attitudes and beliefs of today's workforce will shape future norms of interaction and learning, it is vital to understand how knowledge workers develop GenAI competencies in practice. \rev{Specifically, we explore perspectives that elaborate and contextualise workers' beliefs and learning strategies for GenAI competencies, focusing in particular on extending understanding of overlooked socio-technical competencies (see Table \ref{tab:frameworks_tbl})}. We aim to address two main Research Questions (RQs): 

\begin{enumerate}
     \item[\textbf{RQ\textsubscript{1}}:] What competencies do knowledge workers value when using GenAI tools in the workplace?  
     \item[\textbf{RQ\textsubscript{2}}:]Which strategies do knowledge workers consider essential to developing these competencies?  
\end{enumerate}

In answering these questions, we make three main contributions:

\begin{enumerate}
    \item We conducted semi-structured interviews with 19 knowledge workers \edit{from diverse professions, allowing us to develop an empirically grounded account of how GenAI literacy is practised and negotiated within our sample.} 
    \item We showed that, while workers benefitted from colleagues sharing knowledge, the act of critiquing and obscuring their own GenAI usage was also valuable to them. Importantly, while prior work has interpreted acts of GenAI hiding as rooted in shame \cite{zhang2025}, or a desire to elevate one's status \cite{sarkar2025}, we found that users can also view successful erasure of telltale signs of GenAI outputs as a positive indication of one's professional expertise. However, these behaviours have negative impacts on further opportunities to learn via knowledge sharing, and promote cultures that lack transparency (Section \ref{sec:results}).  
    \item We extend existing AI and GenAI literacy frameworks by highlighting the need to consider human stakeholders both as a resource for learning and as a source of social influence on usage. Building on this perspective, we highlight implications for HCI design to support open dialogue, disclosure, and sustainable knowledge sharing in the workplace (Section \ref{sec:discussion}).  
\end{enumerate}

%% file: 02a_GenAI_in_Knowledge_Work.tex
\section{Related Work}

Our work builds on several lines of research that may be grouped into three areas: \emph{(1)} GenAI in knowledge work (Section \ref{subsec:rw_1}); \emph{(2)} developing digital skills in the workplace (Section \ref{subsec:rw_2}); and \emph{(3)} defining AI and GenAI literacy (Section \ref{subsec:rw_3}). 

\subsection{GenAI in Knowledge Work}
\label{subsec:rw_1}

Knowledge work describes the complex and non-routine labour that requires tailored problem-solving skills and high levels of expertise~\cite{davenport1996, drucker1993, surawski2019}. Knowledge workers in various sectors (e.g., researchers, consultants, creatives, IT professionals~\cite{desordi2021, surawski2019}) comprise a significant part of the world economy. In recent years, there has been an increase in interest and uptake in the use of GenAI tools to knowledge work \cite{eloundou2023, felten2023, chui2023, hatzius2023}. We define GenAI as end-user tools that draw on deep learning approaches and large amounts of training data to create outputs in response to user prompts \cite{chui2023, bommasani2022, sarkar2023}.  
The integration of GenAI tools into knowledge work is expected to bring benefits to productivity, but also challenges \cite{shao2025}. GenAI tools can conduct many core components of knowledge work tasks \cite{brachman2024, retkowsky2024, tomlinson2025}, such as searching and collating information \cite{siu2025, yen2024, zhou2024}, communication \cite{su2023, akdilek2024}, or generating tailored solutions in problem-solving \cite{vaithilingam2022, boussioux2024, sakib2024}. However, recent work suggests that the productivity gains from automating these tasks may be offset by demands for new skill-sets which arise from using GenAI~\cite{constantinides2025a}. Multiple studies have highlighted that the use of GenAI tools presents a shift in users' labour from production to supervision \cite{woodruff2024, simkute2024, lee2025a}. This shift introduces new tasks to pre-existing workflows, such as refining prompts or adapting outputs (i.e., task stewardship) \cite{dang2022, korzynski2023, zamfirescu-pereira2023, lee2025a}, but also require users to make informed decisions as to when and what types of tasks could be delegated to GenAI \cite{jin2024a, kobiella2024}, termed `critical integration' \cite{sarkar2023b}. However, this introduces new challenges. Tankelevitch et al. \cite{tankelevitch2024} highlight how adoption of GenAI introduces significant metacognitive effort by requiring users to clearly articulate and refine their goals when prompting \cite{korzynski2023, zamfirescu-pereira2023}. Reliance on low quality generated GenAI outputs can also threaten the need for human expertise and risk deskilling industries, \edit{negatively impact critical thinking and cognition \cite{lee2025a}}, or promote practice that lose the human characteristics (e.g., empathy) for interpersonal work \cite{woodruff2024, kobiella2024}. As GenAI reshapes the skills and cognitive demands placed on knowledge workers, it becomes increasingly important to re-evaluate which competencies and learning strategies users prioritise in these evolving work environments, and their motivations for doing so. 



%% file: 02b_Digital_skills_in_workplace.tex
\subsection{Developing Technical Skills in the Workplace}
\label{subsec:rw_2}
While traditional educational models in school settings are characterised by structured, theoretical progression and clearly defined performance indicators, learning in the workplace is often more participatory, \edit{and dependent on communities of practice \cite{wenger2000}}, where each worker is expected to play an active role in creating and sharing knowledge \cite{freire2014, tynjala2008}. The social context and priorities of the workplace play an important role in how learning takes place. 

Previous work exploring the development of technical skills (e.g., software use, end-user programming) in the workplace highlights that learning is often shaped by external factors such as time pressures, competing priorities, and a pressure to maintain productivity \cite{lemmetty2020a, shahlaei2022}. As a result, workers tend to rely on strategies that are perceived as readily available and directly relevant, such as informal help-seeking, knowledge sharing with colleagues, or turning to search engines \cite{kiani2020, lemmetty2020a, twidale2005}. Furthermore, an individual's technological competence is often closely intertwined with their professional identity and value \cite{nouwens2018, sarkar2023a}. As a result, users' personal interests and values regarding what technical skills needed are also influenced by the social norms and values of the environment \cite{orlikowski1992, orlikowski1993, orlikowski2000, heyder2021, nouwens2018}. For example, recent studies of spreadsheet users highlight that users may alter the functionalities (e.g., formula-writing \emph{vs.} using Macros) or visual layout of a shared spreadsheet to conform to the needs of the recipient and to avoid incurring negative or unprofessional social cues (e.g., messy presentations) that may be evoked by the way the spreadsheet is presented \cite{xia2025, srinivasaragavan2021, chalhoub2022}. 

Concerns about social scrutiny and judgement based on one's technical competence has been a pervasive barrier to workplace knowledge sharing in a software learning context \cite{giannisakis2022, pipek2012}, but have become even more relevant in the context of GenAI due to trends of stigmatisation against GenAI users. Recent works have highlighted that disclosure of GenAI use is often accompanied with undesirable social consequences, such as lower perceived credibility, quality of work, and lowered trust \cite{schilke2025, rae2024}. This social environment has led some researchers to explore the issue of AI hiding and shame, a behaviour whereby individuals downplay or \edit{intentionally remove evidence} of AI use to avoid the associated social penalty incurred through admitting its usage \cite{sarkar2025, zhang2025, li2024}. For example, once people became aware that GenAI outputs were often biased with specific word choices (e.g., `delve'), there were notable decreases in the word's occurrence as users sought to disassociate their work from GenAI use \cite{geng2025}. However, while AI hiding is seen as a behaviour that counters the overall goal of establishing transparency in responsible GenAI usage \cite{schilke2025}, its impact on the knowledge sharing dynamics that are important for technical skill development in the workplace is not well understood. As knowledge sharing is a key method for retaining and maintaining organisational knowledge and technical skillsets among employees long-term \cite{smith2017, lemmetty2020a, lemmetty2020, collin2021}, understanding the impacts of AI hiding through the lens of knowledge sharing can have important implications for considering how AI literacy could be sustainably supported over time.

%% file: 02c1_GenAI_Literacy.tex
\subsection{Defining AI and GenAI Literacy}
\label{subsec:rw_3}

The traditional concept of `literacy' is about ensuring that members of the population are equipped with the foundational skills needed to understand, express, and communicate using language. In recent years, there has been growing interest across the fields of computer education and HCI in exploring the concept of literacy in relation to AI technologies \cite{cox2024, almatrafi2024}, including but extending beyond GenAI to include robotics, machine learning etc. \cite{long2020, pinski2024}. This research is largely divided into two strands. The first, rooted in the field of education, primarily focuses on defining or measuring AI literacy \cite{long2020, pinski2024, cetindamar2024, heyder2021, weber2023}. The second, which has been of greater focus in the HCI community, is interested predominantly in supporting the development of AI literacy across a diverse range of users \cite{cao2025, xie2025, kotturi2024, ko2025, kaspersen2024}. More recently, works have also drawn distinctions between AI and GenAI literacy. This is due to the widespread popularity and accessibility of GenAI tools among non-specialist users, the need for more specific skills such as prompt optimisation when interacting with GenAI tools, as well as the cross-domain applicability of GenAI tools requiring more careful consideration of how and when such technology should be applied in different contexts \cite{annapureddy2025, liu2025}. 

We observe three main dimensions in which proposed competencies may be clustered: \emph{1}) technological competencies, describing the technical knowledge needed to understand and interact with AI; \emph{2)} learning and reflective competencies, which covers more general metacognitive and thinking skills of reflection, evaluation, and communication; and \emph{3)} socio-technical competencies, which involves understanding the role of human stakeholders throughout the process of AI use. Table \ref{tab:frameworks_tbl} summarises AI and GenAI literacy frameworks across the three dimensions, and highlights how each defines competencies for adult and general audiences. Due to the limited number of GenAI literacy frameworks and the significant overlaps between GenAI and AI literacy frameworks, we review both categories of framework together, and annotate competencies that are specific to GenAI. 

\begin{table*}[ht!]
    \centering
    \caption{List of technological, socio-technical, and learning and reflective dimensions with corresponding interpretations of competencies across AI and GenAI literacy frameworks.}
    \Description{Table listing the technological, socio-technical, and learning and reflective dimensions in the left column, interpretations of competencies within each dimension, and list of references for each interpretation.}
    \label{tab:frameworks_tbl}
    \scalebox{0.99}{
    \begin{tabular}{p{3.5cm} p{7.5cm} p{3cm}}
        \toprule
        \textbf{Dimension} & \textbf{Interpretation} & \textbf{References} \\
        \midrule
        \multirow{4}{=}{\textbf{Technological competencies}} 
            & Understanding what AI/GenAI is, what it is capable of, how it works, and its limitations 
            & \cite{long2020, ng2021, heyder2021, cetindamar2024, carolus2023, pinski2023, pinski2024, cox2024, lee2024a, annapureddy2025, liu2025a}\\
        & Ability to apply AI tools to solve problems, complete tasks, and enhance productivity and work outputs
            & \cite{long2020, ng2021, yi2021, heyder2021, cetindamar2024, carolus2023, pinski2023, pinski2024, cox2024, lee2024a, annapureddy2025, liu2025a}\\
        & Creating and designing AI models, algorithms, and tools
            & \cite{carolus2023, pinski2023, annapureddy2025}\\
        & (\textit{GenAI-specific}) Skill in prompt engineering, articulation and proficiency in communication with GenAI tools
            & \cite{lee2024a, annapureddy2025, liu2025a}\\
        \midrule

        \multirow{8}{=}{\textbf{Learning and reflective competencies}} 
            & Understanding how AI may be perceived, and its impact on users' experiences, social interactions with AI, how accessible and learnable AI may appear to be 
            & \cite{long2020}\\
        & Critical thinking and evaluation of the outputs of AI or GenAI models 
            & \cite{ng2021, pinski2024, lee2024a, liu2025a}\\
        & Metacognitive ability to reflect on, identify, address and manage one's own needs (e.g., what information is needed, how to access it); and how external factors (e.g., AI) may influence the thinking process
            & \cite{yi2021, heyder2021, markauskaite2022, carolus2023, pinski2024, cox2024}\\
        
         & Ability to think creatively about using AI, to translate innovative approaches into practical outcomes and ideas 
            & \cite{heyder2021, markauskaite2022, lee2024a, liu2025a}\\

         & Ability to engage in continuous learning to adapt to developments in the technology
            & \cite{cetindamar2024, markauskaite2022, carolus2023, pinski2024, annapureddy2025}\\

         & Foundational literacy skills in reading, writing, arithmetic; basic understanding of how to understand and communicate through symbols, text, and numbers 
            & \cite{yi2021}\\

         & Regulation of emotions when interacting with AI 
            & \cite{carolus2023}\\

         & (\textit{GenAI-specific}) Ability to detect GenAI-generated content 
            & \cite{annapureddy2025}\\
         \midrule

         \multirow{3}{=}{\textbf{Socio-technical competencies}} 
            & Consideration of the ethical and legal implications of AI use e.g., equality, privacy, transparency, accountability, impact on employment, environmental impacts, trust, safety
            & \cite{long2020, ng2021, heyder2021, carolus2023, pinski2023, pinski2024, cox2024, lee2024a, annapureddy2025, liu2025a} \\
        & Understanding the role of human actors in human-AI collaboration, including human-specific advantages and disadvantages 
            & \cite{cetindamar2024, markauskaite2022, pinski2023, pinski2024}\\
        & Understanding the AI impact on social norms, relations, and collaboration with other human stakeholders
            & \cite{heyder2021, markauskaite2022, pinski2024} \\
        
        \bottomrule
    \end{tabular}
    }
\end{table*}


While socio-technical competencies play an important role for AI literacy, there is a significant focus on educating users on normative ethical implications of GenAI use. However, as stated in Section \ref{subsec:rw_2}, users' personal interests and values regarding different technical skills are closely shaped by their social norms, organisational contexts, and a desire to support the presentation of their own professional identity \cite{nouwens2018, orlikowski1993}. Additionally, only a few frameworks identify the need to navigate social and collaborative contexts in AI use \cite{markauskaite2022, heyder2021, pinski2024}, and they often overlook how such dynamics shape sustained learning.
\smallskip


In summary, our review highlights that our understanding of the competencies needed to use GenAI successfully in knowledge work is still evolving. We also find that current AI literacy frameworks largely overlook the social dynamics of AI shaming, disclosure, and hiding, and in particular its implications for unstructured learning within the workplace. We aim to fill this gap by providing a deeper exploration of the beliefs and experiences of knowledge workers who are continuously learning to use GenAI tools while managing the social dynamics of the workplace.

%% file: 03_Method.tex
\section{Method}
\label{sec:method}
\subsection{Participants}
During participant recruitment, we first identified sectors, which, based on previous economic analyses, were adopting generative AI tools at rapid rates \cite{chui2023, felten2023, singla2025, mayer2025}. We then \edit{identified three core characteristics of `knowledge work' and recruited participants belonging to sectors that met those characteristics. We recruited} from sectors where expertise is developed primarily through formal education or training \cite{drucker1993} \edit{(i.e., IT, Data Science, Law)}, \edit{sectors where tasks involved} non-routine or creative problem-solving \edit{that require} contextual sensitivity to meet stakeholders' needs \cite{davenport1996} \edit{(i.e., Management and Human Resources, Creative industries),  and sectors which} specialised in \edit{knowledge creation and dissemination} \cite{desordi2021} \edit{(i.e., Research and Development). Incorporating diverse perspectives from different sectors is important for ensuring rigour in qualitative research, as our goal is not to produce a generalisable account of human behaviour, but to ensure that we extend existing interpretations of the research subject \cite{braun2021}.}

In order to better access participants from these different occupational roles, we used the online recruitment platform Prolific. Prior evaluations has demonstrated that data quality from participants on Prolific, evaluated through user attention, comprehension, honesty, and care with answering questions, was consistently higher than comparable platforms \cite{douglas2023, peer2022, albert2023}. One of the advantages of using Prolific was that it allowed finer control over participants' demographics and attributes, allowing recruitment to be targeted. We used both Prolific's own in-built filters, based on participants' self-reported demographics when registering for a Prolific account, as well as our own screening survey \edit{to ensure participants met our recruitment criteria, namely, participants should: have occupations belonging to one of the sectors identified above;} use GenAI tools at least once a week and primarily for work-related (as opposed to leisure or personal) purposes; be employed in a full- or part-time capacity in their specified sector and role; have more than 1 year of experience in their current role; and are not working in organisations which explicitly prohibited the use of GenAI tools in their work, as this may constrain the types of learning resources they can \edit{access, which was an important aspect of the research question}. However, we allowed participants to participate if their organisation had no clear GenAI policies, or if participants themselves were uncertain of the policies, \edit{as this allowed us to explore how our participants navigated and developed their own understanding of GenAI competencies independent of top-down organisational guidance}. These criteria ensured that our participants had sufficient familiarity with their specific work environment, domain, and \edit{had the opportunity to access} the relevant GenAI tools and \edit{learning resources} to be able to discuss their experiences, perceptions, and learning strategies in the interview. 

We \edit{initially identified 25 eligible participants for interviews. Six participants (not included in the table) were excluded. Four dropped out of the study before the interview or experienced technical issues during the interview, and two were excluded post-interview because they used tools which we could not reliably verify as using GenAI models. Our final data-set consisted of} semi-structured interviews with 19 knowledge workers \edit{collected from} December \edit{4th} 2024 to  January \edit{17th} 2025, \edit{which meets the standard sample size expected for the field \cite{caine2016a}.} 

Table \ref{tab:p_demog} outlines the demographics of the participants, their job title, organisation size, and experience. Table \ref{tab:p_tools} outlines the types of GenAI tools used by each participant, and the types of tasks which the GenAI tools are used for. 

\begin{table*}
  \centering

  \caption{Demographics of the 19 interview participants, including the participant's ID, age range and gender, job context, organisation size, and domain experience. Participants represented diverse sectors such as academia, law, IT, creative industries, finance, and human resources, with experience levels ranging from 1–4 years to over 10 years. \edit{Participant IDs are not in order as six initially eligible participants were excluded from the final data-set.}}
  \Description{Table summarising the demographics of 19 interview participants, with five columns representing the participant ID, age and gender of the participant, their job context, organisation size, and their domain experience in years. Participant IDs are not in order as six initially eligible participants were excluded from the final data-set.}
  \label{tab:p_demog}

  {\footnotesize
  \begin{tabular}{p{0.75cm} p{1.5cm} p{4.5cm} p{3cm} c}

  \hline
  \textbf{ID} & \textbf{Age (Gender)} & \textbf{Job Context} & \textbf{Organisation Size} & \textbf{Domain Experience}\\
  \hline

  P1 & 45-54(M) & Astrophysicist and academic researcher &  Large (250+ employees) & 1-4 years\\
  P2 & 25-34(F) & Bid writer in marketing agency & Small (10-50 employees) & 1-4 years\\
  P4 & 25-34(M) & Social media marketing and communications manager & Large (50-250 employees) & 1-4 years\\
  P6 & 45-54(F) & Data analyst in mental health charity& Large (250+ employees) & 5-9 years\\
  P7 & 18-24(M) & Data analyst in insurance company & Large (250+ employees) & 1-4 years\\
  P9 & 35-44(M) & Creative copywriter for consumer technology company & Large (250+ employees) & 5-9 years\\
  P10 & 25-34(F) & Cybersecurity engineer & Medium (50-250 employees) & 5-9 years\\
  P12 & 55-64(M) & Software and solutions architect specialising in medical systems & Large (250+ employees) & 10+ years \\
  P13 & 55-64(M) & University lecturer and researcher & Large (250+ employees) & 10+ years\\
  P14 & 25-34(M) & Social media manager and assistant HR & Small (10-50 employees) & 5-9 years \\
  P15 & 45-54(F) & Research professor in digital business & Large (250+ employees) & 10+ years\\
  P17 & 25-34(F) & Trainee solicitor & Medium (50-250 employees) & 1-4 years \\
  P18 & 45-54(F) & HR officer in agricultural institution & Small (10-50 employees) & 10+ years \\
  P19 & 25-34(M) & Sales coordinator, analyst, and distributor for a transport logistics company & Small (10-50 employees) & 1-4 years\\
  P20 & 25-34(F) & Senior accountant & Medium (50-250 employees) & 1-4 years \\
  P21 & 25-34(M) & Software engineer & Medium (50-250 employees) & 1-4 years \\
  P22 & 25-34(M) & Business relationships manager & Large (250+ employees) & 5-9 years\\
  P23 & 25-34(M) & Senior architect & Medium (50-250 employees) & 1-4 years\\
  P25 & 25-34(F) & Marketing coordinator for a financial software company & Medium (50-250 employees) & 5-9 years \\

  \hline
  \end{tabular}
  }
  
\end{table*}

\begin{table*}
  \centering

  \caption{Participants' reported use of GenAI tools and their use cases in the workplace. Tools included ChatGPT, Copilot, Claude, Gemini, DALL-E, and others. These tools were used for a range of tasks such as content generation, coding, debugging, writing support, design, communication, and knowledge management.}
  \Description{Table summarising participants' use of GenAI tools and applications, with three columns showing the participant ID, the type of GenAI tool(s) used by participants, and the corresponding GenAI use cases. Participant IDs are not in order as six initially eligible participants were excluded from the final data-set.}
  \label{tab:p_tools}
  \renewcommand{\arraystretch}{1.3}
  
  {\footnotesize
  \begin{tabular}{p{0.75cm}   
    p{4cm}     
    p{9cm} 
  }

  \hline
  \textbf{ID} & \textbf{GenAI Tool(s)} & \textbf{GenAI Use Case}\\
  \hline

  P1 & ChatGPT & Generating codes for writing software \\
  P2 & Notion AI, ChatGPT & Generating initial drafts and refining writing \\
  P4 & ChatGPT, Quillbot, Gemini & Generating content and ideas for social media, refining writing \\
  P6 & ChatGPT & Generating Visual Basic Advance code and Excel formulas, providing support and solutions to problems in analysis \\
  P7 & ChatGPT, PowerBI/Excel Copilot plugin & Providing advice for debugging, structuring data, performing data analysis according to parameters and analysis goals \\
  P9 & Copilot, ChatGPT & Generating ideas for content, image generation to support communication with designers \\
  P10 & Claude AI, ChatGPT & Generating initial drafts of documentation and policy reports (e.g., exercises to support training on information security)\\
  P12 & Copilot & Answering questions, generating code, providing support for debugging, error checking and conducting unit tests on codes \\
  P13 & Gemini, Copilot, ChatGPT & Generating lecture slides, generating summaries of papers for writing papers, writing emails, refining and editing writing \\
  P14 & ChatGPT, DALL-E, Amper Music & Generating content for social media, generating code for building websites \\
  P15 & ChatGPT & Generating titles for papers, drafting emails, reviewing and generating content ideas for lecture courses \\
  P17 & Lexis+ AI & Answering queries and providing support based on precedent case files \\
  P18 & Gemini, ChatGPT & Drafting communications and emails, supporting user learning by summarising and teaching agricultural terms when writing investor reports \\
  P19 & Claude AI, Lovable.dev, ChatGPT & Creating solutions to support journey mapping and logistics for arranging transport deliveries, generating code to create web platforms for clients, debugging code \\
  P20 & ChatGPT, Copilot & Answering queries and providing reminders of accountancy principles to support auditing process \\
  P21 & ChatGPT, Gemini, Perplexity, Copilot & Support interpretation of legacy SQL code, conduct online search to return documentation of code libraries, writing emails \\
  P22 & ChatGPT, myKnowledge & Answering work-related queries to help with responding to customer queries, proposing solutions \\
  P23 & ChatGPT, DALL-E, Claude AI, Gemini & Generating interior designs in different styles, generating Powerpoints for work presentations \\
  P25 & Gemini, ChatGPT & Proofreading writing and content for blogs, summarising concepts to support learning about search engine optimisation and marketing-related skills  \\

  \hline
  \end{tabular}
  }
  
\end{table*}

\subsection{Interview Protocol and Data Collection}
We developed a semi-structured interview protocol to facilitate data collection. As our participant sample spanned a wide range of sectors, we ensured mutual understanding by establishing at the beginning of interviews: \emph{(1)} the occupation and professional responsibilities of the participant; \emph{2)} their understanding of the term `generative AI'; and \emph{3)} how participants currently use GenAI tools in their work. This provided opportunities for the interviewer to ask for clarifications regarding participants' backgrounds, informally evaluate participants' technical expertise and to adjust their terminology appropriately to ensure communication, and to guide conversations to focus on GenAI tools if they struggled to distinguish between similar technological products (e.g., generative \emph{vs.} rule-based Chatbots). 

During interviews, we prioritised flexibility, open questions, and following up on participants' statements in order to better capture participants' individual perspectives, experiences, and beliefs. Prior to data collection, the protocol was also piloted with two participants (one PhD researcher, one creative copywriter) to evaluate the terminology used and to ensure that the questions asked prompted relevant interpretations and discussion. The main topics of inquiry explored in the study involved: 

\begin{itemize}
    \item What competencies do participants perceive to be most important to using GenAI in their work? 
    \item How did participants develop their current practices and workflows for using GenAI tools? 
    \item What resources or strategies do participants use to help them learn to use GenAI tools, and why? 
    \item How do users communicate about their skills and knowledge of GenAI tools within their workplace? 
\end{itemize}

Interviews were conducted online by the first author, using Microsoft Teams. Eligible participants were identified via the initial recruitment survey distributed on Prolific, and invited to book a suitable timeslot for the interview. The interviews recorded and transcribed using Microsoft Teams' built-in functionality. Transcriptions were then manually corrected in accordance to video recordings. Interviews typically lasted between 50-60 minutes. Ethics for conducting the study was obtained and approved by the University College London internal ethics committee. All participants were informed of the purpose of the study and consented to participate. \edit{To accommodate for participants' needs \cite{alkhatib2017}, participants were invited to book timeslots for the interview according to their own availability. All participants were compensated £0.70 for completing the 2-minute screening survey, regardless of whether or not they also completed the interview. All interview participants, including those whose data were ultimately excluded post-data collection, were compensated at £13 an hour, proportional to the amount of time taken in their respective interviews.} 

\subsection{Analysis}
\subsubsection{Positionality}
\edit{Our research team consisted of five members with expertise in research with Human-Computer Interaction, responsible GenAI, software learnability, and technology use in the workplace. Our overall approach to collecting, analysing, and presenting the data is rooted in} critical realism \cite{braun2019, pilgrim2014}, meaning we recognise that both our interpretation as well as participants' own articulations reflect their subjective experiences and beliefs about GenAI competencies and the resources necessary to support their learning \cite{hall1997}, \edit{rather than reflecting an objective or concrete reality}.\edit{We adopt this position as it allows us to focus explicitly on understanding} participants' personal experiences and values. This is important for \edit{addressing the primary goals of this paper, which is to} develop alternative ways of thinking, clarifying, and extending existing frameworks of AI literacy and competencies. 

\edit{Due to the nature of critical realism, it is also important for transparency and rigour to disclose our authorial positionality. The first author, who collected the data and led the analysis process, is an academic researcher with a primary interest in organisational learning. They are guided by the view that learning is an inherently social process, shaped and developed through interactions with a wider community} \cite{wenger2000}. \edit{As such, our research process focused on} relevant competencies in GenAI, considering not only the technical knowledge of GenAI interactions or understanding of GenAI models, but also on the skills surrounding GenAI use that are important for users' social outcomes, such as reputation and maintaining positive relationships \cite{ackerman2000, weber2023, pinski2024, heyder2021}. \edit{To support reflexivity and critical reflection, the analysis process was supported by two co-authors with expertise in GenAI research, who offered alternative perspectives for the first author. Two additional co-authors with expertise in workplace research further provided reflections over the main themes during write-up. In addition, the first author, who was the only one to interact directly with participants, had no prior experience with any non-academic sectors participants were affiliated with. This allowed for consistency during data collection when questioning and establishing an understanding of participants’ work contexts and their GenAI usage.}

\subsubsection{Thematic analysis}
We conducted thematic analysis using a template approach \cite{king1998}. An initial framework, i.e., a `template', is applied to scaffold the initial coding and interpretation of the data, but new categories may be added or merged \note{during coding} to reflect researchers' evolving understanding of the data. This method is particularly suitable for identifying nuances where interpretations of the data diverge from pre-existing frameworks. While we acknowledge that there is no widely agreed definition of the competencies necessary to become AI literate, and that the importance of different competencies vary depending on researchers' domain and pedagogy \cite{almatrafi2024, markauskaite2022}, the aim of our analysis was to provide insights into the underlying reasoning and values behind how knowledge workers conceptualise and develop GenAI competencies. 

We \note{applied} different theoretical frameworks to organise the initial analysis of our data, corresponding to \note{each} research questions. For RQ1, we used two AI literacy frameworks to provide an initial \note{mapping of} how participants conceptualised GenAI competencies. The first framework was developed by Pinski and Benlian \cite{pinski2024}, and was chosen because: \emph{(1)} it specifies competencies shaped by AI's unique features (e.g., inscrutability, autonomy, and its ability to learn) which challenge the assumptions of conventional computer learning; \emph{(2)} it provides a socio-technical perspective that goes beyond technical knowledge to include the social skills needed to navigate AI use with other stakeholders; and \emph{(3)} it was developed \note{based} on a recent \note{literature review of AI literacy}. The second framework was the GenAI literacy framework proposed by Annapureddy et al. \cite{annapureddy2025}, which, to our knowledge, is one of the only attempts to date to provide a systematic overview of the competencies unique to GenAI. Related categories from both frameworks were merged together for the initial template. 

To address RQ2 and understand the general facilitators and barriers which affect users' choice of learning resources and strategies, we used the COM-B model of behaviour to form the initial template for analysis. This model suggests that observable behaviours (in our case, strategies for learning such as trial-and-error) manifest when three core components of behaviour align – capability, opportunity, and motivation. Each component can also influence the other \cite{michie2011}. Though typically applied to help researchers map out and develop their understanding of a single behaviour with the goal of providing prescriptive recommendations, we applied this model to provide a comprehensive overview of \note{the external and internal factors driving users' choices of GenAI learning strategies.}

The first author conducted open coding over the interview data-sets, which were then organised according to the three theoretical frameworks outlined above. Particular focus was placed on categories where data diverged from, clarified, or identified areas of overlap, between the two AI literacy frameworks used. We established rigour in the analysis by ensuring that the first author engaged in principles of reflexivity, and by incorporating the perspectives of the second and third authors throughout analysis \cite{braun2019}. \note{Both the second and third author regularly reviewed quotes, code, and theme descriptions as they were being developed. This} involved interrogating the overall clarity of themes, the first authors' interpretations of the relationships between contributing factors, and the evidence selected to support claims. This ensured critical reflection throughout the coding and writing process via multiple alternative perspectives.

%% file: 04_2_Results.tex
\section{Results}
\label{sec:results}

In answer to RQ1, we identified two core competencies which our participants described as important for using GenAI in practice: \emph{(1)} \update{a technological competency, namely, the} awareness of potential GenAI use cases \edit{(see Section  \ref{sec:comp_awareness})}; and \emph{(2)} \update{a socio-technical competency,  namely,} the ability to \edit{remove cues} of GenAI usage \edit{via concealment and critique (detailed in Section \ref{subsec:comp_removing_cues}}) \update{as a means to manage quality of work, as well as one's own presentation and image to others}. In answer to RQ2, we found that these competencies are developed through two corresponding learning strategies: \emph{(1)} learning from others was vital for developing awareness of potential GenAI use cases \edit{(see Section \ref{sec:learn_from_others})}; and \emph{(2)} learning through hands-on experience with GenAI tools played an important role in developing users' ability to detect and mask evidence of GenAI usage, as they become more accustomed to biases and limitations in the GenAI outputs \edit{(see Section \ref{sec:hands_on})}. 

\begin{figure*}
    \centering
    \includegraphics[width=1\linewidth]{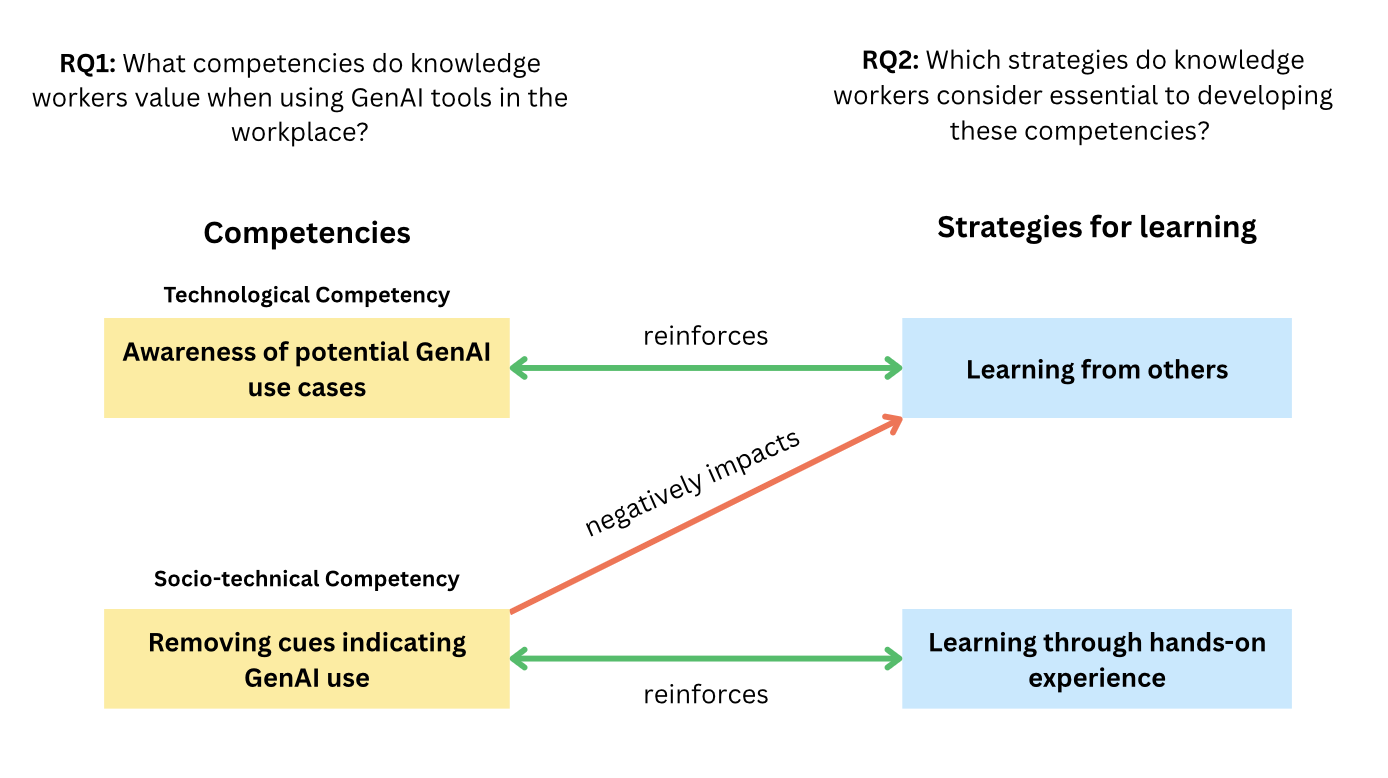}
    \caption{Relationships between competencies and learning strategies identified in the study. Knowledge workers valued awareness of potential GenAI applications and the ability to \edit{remove cues indicating GenAI use (detailed in Section \ref{subsec:comp_removing_cues})} These competencies were reinforced by two key learning strategies (i.e., learning from others and hands-on experience), while \edit{removing cues} negatively impacted opportunities for shared learning.}
    \Description{Diagram representing the relationships between competencies and learning strategies identified in the study, depicted through four colour-coded elements and arrows to denote relationships between elements. Two competencies (`Awareness of GenAI use cases', \edit{Removing cues indicating} GenAI use') are depicted as yellow elements, and two strategies for learning ((`Learning from others', `Learning from hands-on experience') are depicted as blue elements. Elements are connected by arrows denoting that `Awareness' and `Learning from others' reinforce each other, while `Removing cues' and `Learning from hands-on experience' reinforce each other, and `Removing cues' also negatively impacts `Learning from others'.}
    \label{fig:results}
\end{figure*}

Taken together, these findings reveal a central tension. While \update{the development of technological competencies, in particular developing awareness of potential AI use cases, benefit from social processes, knowledge exchange, and learning from others, individuals exercising socio-technical competences and considering the social impacts of GenAI use often engage in prevalent practices} of concealing and critiquing GenAI usage in ways which constrain opportunities for open knowledge sharing. Importantly, concealment was not always framed negatively as shame. For some, successfully erasing traces of GenAI use was a positive signal of professional expertise. Figure \ref{fig:results} summarises the main themes for each of research question, how they relate to each other, and the contributing factors defining each theme. 

\subsection{Technological Competency: Awareness of Potential GenAI Use Cases}
\label{sec:comp_awareness}
Discoveries of novel GenAI tools, use cases, and methods of interactions were often described as key moments which redefined participants' understanding of GenAI technology. P4, describing how her awareness developed over time, stated: "\textit{Before, I just used [ChatGPT] for, let's say, text, right? [...] And I found out, OK, you can use it to generate images, you can use it actually to do analysis.}" Awareness helped people feel like they could better understand the impact of GenAI tools, what it meant for the areas of work or the types of tasks that they were interested in, and helped them to access new possibilities for improving their current workflow and practices involving GenAI.  

\subsubsection{Awareness \edit{provides a competitive advantage}}
Maintaining awareness of GenAI's potential use cases was important to participants because GenAI tools possessed potential utility across a variety of domains. For participants such as P14, it was empowering to feel that they may be able to harness such technology for their own purposes: "\textit{I know virtually everything now, can be done [...]. You can solve any, most of every problem now, with AI [...]. The next thing was me learning myself. That's what puts me on top of my game."} For some participants, the scope of GenAI's potential functionalities included not only the technology's current capabilities, but also its anticipated future developments. P19, who closely followed GenAI industry leaders on the social media platform X, explained: "\textit{If [industry leaders] say that you['re] build[ing] Sora [video generating AI tool], I'm trying to understand what would be the end goal of Sora? Is it something that could actually replace the movies that we watch?}" Participants were motivated to keep up with GenAI developments in order to harness it more quickly and address their current problems, or simply to reduce the perceived threat to their own job and career security by attempting to predict future industry trends. P10 shared: "\textit{If care is not taken, children who are 15 years old now [...] they might know how to use this tool, AI, more effectively and efficiently than I do, and before I realise, that could also be a threat to my own job or my career.}"

However, not all participants were motivated to learn due to career concerns. This was particularly the case when GenAI tools were applied and packaged to users as tools serving very specific or narrow functionalities. One example is P22, whose usage of GenAI primarily consisted of an intelligent internal search engine and knowledge management tool recently implemented in their organisation. He explained: "\textit{For learning about something that's already implemented and you're not going to utilise it further than literally just [submitting queries], you're going to learn a lot of areas which [...] is going to be absolutely pointless [...]. It would be a massive waste of my time.}" 

\subsection{Strategy for Learning: Learning from Others}
\label{sec:learn_from_others}
\subsubsection{Addressing knowledge gaps}
As awareness of potential GenAI use cases was important for \edit{increasing users' competitive advantage as knowledge workers}, participants were wary of knowledge gaps that may arise due to limitations in their own imagination, interests, and access to information. P15 described how recommended posts on LinkedIn, a professional social networking platform, helped her identify new directions for learning: "\textit{On LinkedIn, when I don't look for things, these things are posted, and they'd come to me and I'd become aware of something that I wouldn't be aware of, right?}" P12, a software engineer, explained that while his preferred approach towards adopting new technology is to explore and apply trial-and-error until he obtains an ideal outcome, "\textit{when you go on your own, you learn the bits that you find interesting. There may be gaps in what you do [...] There could be things there that you're missing.}" 

Incorporating the perspectives of others – regardless of whether their experience aligned with users' domains – was seen as a key way of addressing knowledge gaps. Informal, unstructured knowledge sharing in the form of discussions or casual conversations was a common avenue for learning. Several individuals shared their discoveries of new GenAI tools with colleagues, as P23 explained: "\textit{Sometimes I do teach [my colleagues] examples of AI tools I use [...] from their own perspective, they can also tell me that this [tool] is also better.}" For some participants, knowledge sharing during the initial adoption of GenAI tools was naturally incorporated into structured weekly meetings: "\textit{It was just a natural topic of conversation [...] `What are people using it for? How useful do they find it?’ [...] One of the things that came out of [these meetings] was [using Copilot to conduct] the unit test. It hadn't occurred to me to do that, so I went and looked at that further.}" (P12). 

\subsubsection{Accelerate learning and adoption process}
\label{subsubsec:accelerate_learning}
For some individuals collective learning and experimentation helped to accelerate a team's mental model of different GenAI tools: "\textit{If a tool isn't particularly good at something, then you'd just be going around in circles [...] It's just a lot quicker if there's a few of you trying these things out and then sharing that it worked or not.}" (P9) While often, participants cited altruistic motivations for sharing GenAI knowledge, in some cases, users strategically share knowledge with target individuals to normalise GenAI usage, as P18 shared: "\textit{I find myself most of the time encouraging my supervisor to encourage his bosses to incorporate the use of AI [...] Him just giving it a go ahead, or just be willing to try it out, letting us try it out, it'll be good for me.}" 

\subsubsection{Seeking guidance and quality assurance}
In some cases, the inherent uncertainty associated with the use of GenAI – for example, when a use case is novel or niche, or when users are lacking in expertise – seeking guidance and clarity from others can also be important for verifying and ensuring the quality of GenAI outputs. P18 explained how she and her supervisor would collaborate over her use of GenAI, stating: "\textit{My supervisor is not inclined to AI, and I use AI on most of my work activities, so I usually double check with him [...] and the knowledge or the aspect that I'm using AI on. So that's just to make sure it's all correct, and he's on tow with it and that we understand each other}". Similarly, P20, an accountant, shared that: "\textit{Despite getting an answer from AI, you still have to confirm, at times from our senior provider, at times from our colleagues [...] for the reliability of our data.}"  

\subsection{Socio-technical Competency: Removing cues indicating GenAI use}
\label{subsec:comp_removing_cues}

GenAI outputs were often perceived to be different to human outputs. \note{Participants described different cues which they felt indicated evidence of GenAI use. For some, the cues were identifiable in specific patterns of behaviour and word choices.} P21 suggests that GenAI tools display different `habits' to human developers when coding: "\textit{When AI writes the programmes, especially ChatGPT, it adds a lot of comments which normally we developers don't do.}" For others, evidence of GenAI-generated outputs is detected through a perceived lack of contextual understanding or authenticity. For example, P2, whose work involves writing bids for formal tender opportunities, describes how she can tell when her colleagues have "\textit{popped a question in [ChatGPT], it's regurgitated out a piece of writing and then they've sent it over}", because the GenAI tool "\textit{doesn't have that ability to, (1) read between the lines; and (2) just really understand what points to emphasise}." \note{These characteristics were often associated with lower overall quality in work}. P9 described: "\textit{I've seen content where I feel like, I can tell it's AI generated [...] It wasn't written by a professional because it's just a little bit average, if you like. It's just not as clever as it should be.}"

\edit{Developing the competency to remove both types of cues of GenAI use is therefore essential for users. We identified two approaches towards cue removal which we differentiate based on participants' motivations. The first is `concealment', which involves an intentional act of hiding GenAI usage, motivated by social stigma. The second is `critique', where cue removal is primarily motivated by concerns regarding the quality of the work output, and often driven by a negative assessment of GenAI's capacity to meet the needs of the work. These approaches do not appear to be mutually exclusive, but we highlight the difference in motivation between the two as each can achieve different goals for users, and have different implications for users' willingness to disclose their GenAI use. Below, we highlight how `critique' can serve as an act of quality control and a behaviour to signal domain expertise and establish role effectiveness, while `concealment' is predominantly a reaction to perceived social stigma and external judgement.} 

\subsubsection{\edit{Critique as quality control over GenAI outputs}}
\label{subsubsec:critique_qc}
Participants who felt they were able to identify and critique the limitations of GenAI-generated outputs often viewed themselves as maintaining important quality control over these outputs. For P2, \edit{this behaviour extended to her colleagues' work. By} identifying or even gently calling out colleagues for their uncritical usage of GenAI outputs, she could "\textit{make the point that [she] know[s] that they've used it [...] to know that [she] knows they haven't done that piece of work [... and] also to make the point that [she] can pick up on AI quite easily."} These interactions could then be used as examples to highlight why it was important to reflect on appropriate reliance on GenAI: "\textit{When I talk to people about it at work [...] my point is, AI is not the solution. We can utilise it, but it's not going to replace anything we're doing because it's just not at that level yet.}" For others, where authenticity and uniqueness of the content generated is an important aspect of their work, incorporating their own efforts and perspective is an important strategy for mitigating the negative implications of `generic' GenAI outputs. P25, who generates creative content for marketing, explained: "\textit{I just don't feel confident enough, just giving whatever [ChatGPT] gives me to someone else. I always want to make sure that I've looked at it with my own eyes and put my own personal stamp on it, where I feel necessary or applicable.}"

\subsubsection{\edit{Expert identity is validated through GenAI critique}}
\label{subsubsec: domain_expertise}
As GenAI outputs can be viewed as low quality, the opportunity to critique and correct its outputs was also viewed by some participants as opportunity to signal domain expertise \edit{to others (see P2 in Section \ref{subsubsec:critique_qc}) or to oneself.} P19, discussing challenges with debugging when building software using GenAI tools, remarked how it is much easier "\textit{if you have expert knowledge of the particular industry or particular problem that you're trying to solve [...] The AI is a bit confused, [so] I'll jump in and I'll fix it.}" Identifying limitations in GenAI's capacity to address context-specific work can also help to \edit{validate individuals' expertise in relation to GenAI. As P25 explained: "\textit{I can't say I'm the expert in the industry or the expert in marketing, definitely not. But I have a pretty good understanding of how this company works and what people's expectations are [...] ChatGPT, it just probably doesn't have that.}"} P13, a lecturer who uses ChatGPT in various writing tasks, described how he would use the guidelines for detecting GenAI use among his students to monitor his own writing and maintain his personal style: "\textit{There are indicator words, as we call them, that students have been using generative AI, so they're words like `leverage', `delves', `underscores', `deep dives' etc. So when I write, I make sure that those words are not in. [...] So I use [the guidelines], as much as possible, to give my own voice to the information.}" He later remarks that: "\textit{I know people have strategies they use on AI that works really well. If you're very clever, no one actually knows you've used it.}"

In contrast, lack of domain expertise is primarily associated with over-reliance, or lack of critique, towards GenAI outputs. P4, who worked as a social media marketing manager, described his approach towards training younger professionals as the following: "\textit{If you're new, coming up, I'll say: `OK, don't rely on it too much or use it too much. You're trying to develop your own creative skills', [...] but if you've been in this work for a while, of course you can use it.}" On the other hand, P7, a data analyst, assumed that only he and his colleagues would be using GenAI tools on account of their relatively more junior position, and described his surprise when he discovered that his boss also used GenAI in his work: "\textit{I mean, he's the boss. So I thought he just wouldn't use AI, you know, like he would just come up with all these high-level reports and the likes from his head.}" 

\subsubsection{\edit{Concealment as a strategy for managing social stigma}}
Users were motivated to \edit{conceal cues of} GenAI usage due to social stigma and potential negative judgements from those around them. P25 shared her speculation of her colleagues' beliefs: "\textit{There might still be a lot of concern about... is [GenAI] accurate? Is it ethical? [...] I've not come across anyone who actively expresses these views, but you never know.}" For several participants, discussions of GenAI usage was seen to be potentially controversial and likely to incur negative judgement on themselves or their work, and was thus thought to be safer to avoid. P17 described her dynamics with older colleagues: "\textit{I think with the older generation [...] they look down on generative AI [...] they would say, `why don't you look at books and do it manually?' [...] They're quite set in their ways, so it's just best not to agitate them by saying that we use a faster, quicker, more efficient method.}"




\subsubsection{\edit{Removing cues of GenAI use} negatively impacts access to shared knowledge}
\label{subsubsec:negative_impact_KS}

Though the ability to \edit{critique and conceal} GenAI use allowed individuals to maintain quality over GenAI outputs and \edit{validate their own expertise}, it also had a negative impact on others' access to shared knowledge, \edit{and their overall awareness of whether or not GenAI is used}. \edit{Part of the challenge lies in the fact that users differentiate GenAI usage based on familiarity with colleagues' capabilities, styles, and work outputs, as P4 explained: "\textit{If it's someone I know, I know this person's capability, so when they say `This is my own work', you know [if] they used [GenAI]}". He acknowledged, however, that this task of differentiation was "\textit{not easy, because [GenAI] is so advanced.}"} Concerns are exacerbated by environments where GenAI use is not formally acknowledged or established, leading to significant uncertainty and a lack of transparency, even if usage is informally condoned. P10 described: "\textit{I wouldn't want to tell you I use AI for my day-to-day job while we still have these governance and policies and controls in drafts [...] I feel everyone uses AI in one way or another, but no one is saying it to each other because it's not policy enforced yet.}" \edit{In P25's case, it was only through direct observation of colleagues using GenAI tools on their screens that indicates to her that these tools are used in other areas of their work.}


Breaching the subject of others' GenAI use required significant social sensitivity, while disclosing one's own usage required ignoring potential feelings of embarrassment. P2, who would openly question her colleagues for undisclosed GenAI use, explained how careful she is with her tone: "\textit{When other people will do the same job as me [...] and I look at their responses. I'll be like ‘guys, you've just popped this into AI, haven't you?’ Because I find it really obvious. [...] the way I approach it, I'll just be like, jokingly, we just sort of have like a laugh about it.}" On the other hand, P4, when describing his feelings when sharing his use of ChatGPT with others, explained: "\textit{Where I'm working, I have my device connected to a big screen [...] I don't hide to use ChatGPT, right? I don't hide it. I'm not ashamed of using it.}"

However, in other cases, users can refrain from disclosing their experiences with GenAI due to a perceived lack of interest from those around them. This may be shaped by a strong cultural focus on outcomes over processes, which can also stifle discussions around the use of GenAI. P21, a software engineer, summarised his experiences as follows: "\textit{Just get the job done. However you did it, nobody cares. So [...] don't make us talk about it. [...] Maybe it's just indifference, or not the focus for the moment. It's not yet a big deal for us to start adopting it in some ways.}" P6, a data analyst whose work was primarily used to support clinicians, therapists, and other admin staff, similarly expressed: "\textit{They [her colleagues] just want the answer [...] They wouldn't be interested in how I did it. I'm ever so excited if people ever are, but they never are.}" For some individuals, such norms can make even informal conversations around tool use appear irrelevant, or even socially inappropriate. P15, a professor in digital business, explained that, despite experimenting with GenAI for research analysis, and even developing a course to support PhD students in using GenAI for research, outside of these work contexts, informal conversations about GenAI was "\textit{not suitable. It's not interesting. I don't go into a coffee morning and start talking with people about generative AI. People'd think 'are you crazy?'}" She likened the importance of using GenAI to existing software, stating: "\textit{More generally, I don't talk about [GenAI], because I wouldn't talk about how I use Google or Excel or things like that.}" P12, a software engineer who uses GenAI extensively when coding, echoed this sentiment: "\textit{You wouldn't consult your colleague for every line of code that you wrote on your keyboard, for every key press, and I think AI sits at that level}."

The desire to maintain a positive social image can thus negatively impact disclosure and knowledge sharing essential for developing users' technical skills. One key example exemplifying this issue is when P14 described how a colleague asked him to solve a problem she had encountered with Excel formulas. Despite referring to an GenAI tool himself to resolve the bug, P14 chose not to disclose his use of GenAI to his colleague, stating: "\textit{She thought I was a genius. [Asked if he told colleague about using AI] No, no, I did not [tell her]. I think because we didn't talk about how did you get it done. [...] I think that's putting me ahead. Kind of like, they can always come to me [...] like, 'oh, don't worry if you have issues, just walk up to [P14]. He's going to help you out.'"} 

\subsection{Strategy for Learning: Learning through Hands-On Experience}
\label{sec:hands_on}
\subsubsection{GenAI competence as implicit knowledge}
Several participants subscribed to the idea that the skills required to support GenAI use is implicit and challenging to articulate. P9 explained: "\textit{It's like, knowing how literal to be, how short and long or accurate your prompt should be, being able to spot when it's not quite right[...] you develop a sense of, `there's a chance I could have made a mistake in this long block of complex text [prompt]'.}" As a result, some participants, such as P15, suggest that GenAI literacy can only be developed through hands-on experience: "\textit{I think it's not something you can read about, you need to test to use them and to learn how to use them to understand the strengths and weaknesses.}" 

\subsubsection{Contextual evaluation of GenAI tools}
Participants felt it was important to identify the contextual appropriateness of different GenAI tools and outputs. The extent to which GenAI tools can be appropriately relied on, especially in creative domains, can be highly subjective and contextual, and thus users must individually evaluate and compare GenAI outputs with their own domain knowledge to ascertain which tools are most contextually appropriate or useful. For example, P9, who compared the outputs of different GenAI models for copywriting, reflected on the subtle differences across these tools: \edit{"\textit{So we used Copilot, but we also used ChatGPT and [...] Gemini [...], they were all kind of fairly similar, but [...] one might have been better at character counts. One might have been better at [...] creativity and doing funny stuff.}"} P4 similarly described how he might directly test and compare across different models in order to draw conclusions about what was more appropriate for his context: "\textit{Let me explore with Bard, let me explore with Perplexity. Because I think the algorithms are different [across different GenAI tools]. They kind of do the same thing, [but] like little differences [...].}" Because the ability to evaluate GenAI outputs is closely related to how one perceives one's own domain expertise (as seen in Section \ref{subsubsec: domain_expertise}), hands-on interaction, experimentation, and comparison are viewed as important for developing both domain knowledge and GenAI competencies simultaneously. 

\subsubsection{Understanding of technical limitations }
GenAI models were often inconsistent in processing prompts and suffered from various limitations. As P2 outlined: "\textit{I've learnt that you have to word things in a certain way, and like, if you put a couple instructions in one sentence, then you might find that the attention is paid to, like, the first part and not the second.}" Such inconsistencies meant that errors or hallucinations were unexpected, and were often presented plausibly so that it is difficult to users to detect, requiring caution and direct supervision over model outputs. P6, who used ChatGPT to help her write formulas for spreadsheet calculations, shared: "\textit{When I first started using [ChatGPT], I think `God, that's amazing. That came back really quickly and I'm sure that's right,' and I'd do it and it wouldn't be right."} 

\subsubsection{Development of metacognitive understanding}
Finally, users had to develop their own metacognitive understanding of how to prompt and articulate their own needs sufficiently to the GenAI model. P9 shared how interactions with tools such as ChatGPT and Copilot helped him to recognise what specifically needed to be articulated. "\textit{I thought that'd be obvious, but now I realise I need to state everything that I also don't want [...] I need to also state, what I don't want to happen as a kind of spin off result.}" For other participants, direct experience and interactions helped them to recognise the importance of having a clearly defined goal prior to interacting with the GenAI tool. P6, who was initially uncritical of ChatGPT's outputs but whose thinking had evolved over time, stated: "\textit{So I'd have to think about what I want it to look like and what I want it to show me.[...] I know enough to question what it's telling me. I can read [ChatGPT's answer] and think, `well, no, that's not going to work' [...] So then I put another prompt in and it goes backwards and forwards a little bit until we get to where I want it to be.}"

%% file: 05_Discussion.tex
\section{Discussion}
\label{sec:discussion}

The goal of our study was to extend existing interpretations of GenAI literacy by providing detailed insight into the beliefs, learning strategies, and in particular extending understanding of how users interpret socio-technical GenAI competencies within a sample of 19 knowledge workers. We found that, while participants valued learning from their colleagues' experiences in order to expand their own awareness and knowledge, the ability to remove cues indicating their own GenAI usage was also important. \update{This activity inevitably reduces others' ability to detect GenAI usage, and therefore to develop awareness and access to others' experiences. This promoted a lack of transparency around GenAI use which was exacerbated in organisations with unclear policies. However, we also found that the ability to remove obvious GenAI `tells' was not always motivated by stigma and fear of social judgement (though this was an important influence), but could also be interpreted as a positive indication of users' domain expertise. These findings together surface a tension between individual identity work and collective learning that has meaningful implications for how we design for and support GenAI literacy in practice.} 



\update{Our study contributes and extends existing GenAI literacy frameworks by encouraging researchers to reflect more deeply on how competencies for interacting with GenAI manifest for users in context.} In the following section, we reflect on and explore the potential implications of our findings. 

\subsection{Contrasts with Prior GenAI Literacy Frameworks}
\update{Our findings challenge current interpretations about the competency needed to detect GenAI outputs.} The most recent GenAI literacy framework proposed by Annapureddy et al. \cite{annapureddy2025} positions the ability to identify cues of GenAI-generated outputs as a way to shield individuals from being misled by bias, errors, and misinformation perpetrated by other GenAI users. However, our findings in Section \ref{subsec:comp_removing_cues} demonstrate that users apply their knowledge of GenAI cues to remove evidence of GenAI usage in their own work as well. While previous literature have predominantly focused on AI hiding behaviours as an act of shame \cite{zhang2025, sarkar2025}, and indeed, we also observed social stigma as an important motivator for concealing GenAI use \cite{schilke2025, sarkar2025, rae2024}, our study highlights how the ability to remove cues of GenAI use can both serve to improve the perceived quality of work, and help users to validate their own domain expertise over those who `fail' to hide their use of GenAI tools. Indeed, GenAI hiding could even be applied to accumulate personal reputational credit for participants, as we saw with P14, who utilised GenAI to solve a colleague's problem, but took personal credit for developing the solution.
 
 \update{This suggests a potential need for researchers themselves to reframe the behaviours around what is currently perceived as GenAI hiding or GenAI shaming, to consider how such behaviours may in fact benefit other user needs. This mirrors similar research into GenAI non-use, where, rather than viewing non-use as an unambiguously undesirable rejection of a `useful' tool, researchers have instead highlighted how non-use can instead reflcet strategic acts of identity protection on behalf of the user, to preserve essential skills and knowledge rather than delegating to the user \cite{cha2025, woodruff2024}.} Indeed, \update{the term `workslop' has also been applied to describe `obviously' generated GenAI content which is shared without consideration or tailoring for recipients' needs, and so, while we highlight the negative impact of GenAI cue removal on transparency for the purpose of learning, we also recognise that this behaviour can mitigate many of the negative impacts of otherwise producing `workslop', which can damage collaboration, social relations, as well as productivity \cite{niederhoffer2025}.} 

 \update{While maintaining awareness and understanding of possible GenAI applications and interactions is a key competency according to our participants, and also well-emphasised in existing GenAI literacy frameworks \cite{pinski2024, cox2024, annapureddy2025}, our study highlights how the execution of different competencies can impact each other. Namely, we highlight how hiding of cues indicating GenAI usage is viewed as an important competency for quality control and affirming users' domain expertise, but can impact wider engagement and visibility of practices and others' ability to learn from others' experiences. Taken together, these findings suggest that competencies around GenAI use cannot be understood in isolation, and that recommendations to support GenAI literacy in the workplace would benefit from better understanding how employees experience and address GenAI usage in situated contexts \cite{liu2026}.} 

\subsection{The Risks of Non-Disclosure around GenAI Tools}
\label{subsec:disc_risks_non_disclosure}
In our study, participants described concerns with disclosing opinions and experiences of using GenAI to those who may judge them negatively for it, as well as navigating potentially negative social conflicts with other colleagues during discussions, even when personal experience and observations told them that GenAI use was prevalent. More recent works echo these findings, showing that self-reports of GenAI usage were up to 40\% lower compared to reports of others' use of GenAI, reflecting a significant social desirability bias when it comes to disclosure around GenAI use \cite{ling2026}. 

As a practice, GenAI hiding can have important consequences for the long-term sustainability of learning through knowledge sharing and supporting norms of usage. \update{While ultimately an individual decision, hiding behaviours and overall lack of disclosure around GenAI usage can contribute to a collective sensemaking process which normalises `organisational silences' – an absence of discourse, sharing of perspectives and voicing of concerns within an organisation – on GenAI use \cite{morrison2000}. The manifestation of such organisational silences can have significantly negative implications for organisations' ability to develop an accurate understanding of GenAI usage, prevalence, to detect errors, and thus to engage in decision-making and policy-making which can maximise GenAI effectiveness in organisations.} Such patterns have previously been observed in more established software. For example, research in spreadsheet learning highlight that individual perceptions around the social acceptability and reputational gains of discussing a software can have negative implications for their willingness to engage in sharing practices \cite{xia2025, xia2024}. In the spreadsheet context, this lack of communication can lead to significant technical knowledge and skill loss within organisations \update{which pose practical problems when users engage with the artefacts produced through spreadsheet knowledge} \cite{smith2017}. \update{In the GenAI context, however, a lack of discourse and GenAI hiding means individuals' knowledge becomes inherently private and invisible to the organisation \cite{xiao2026}.  This will pose complications for organisations, who may not be able to identify when and where knowledge and productivity is lost, or even what knowledge is lost, nor be able to necessarily replicate such knowledge and the positive outcomes associated with it.}

While other AI literacy frameworks have highlighted the broader need for understanding the use of GenAI in social contexts \cite{pinski2024, annapureddy2025}, \update{by focusing on knowledge workers' individual perspectives, we highlight how users navigate the (lack of) organisational structures and clear social cues around GenAI use by adopting concealment practices. By developing a deeper awareness of individuals' priorities and experiences, we can support educators, researchers, and management to implement structures and interventions which could better address these challenges}. 

\subsection{The Challenges of Understanding GenAI's Potential}
\subsubsection{Lack of discoverability}
\update{Despite the perceived universality and versatility of GenAI tools in a wide variety of contexts and tasks \cite{schellaert2023, annapureddy2025}, translating this potential into practice remains challenging for knowledge workers. The experiences of our participants in Section \ref{sec:learn_from_others} showed that, while maintaining awareness of possible applications is seen as an essential technological competency, the discoverability of capabilities beyond users’ current usage patterns remains limited without external support \cite{mackamul2023}.} Supporting user discoverability is a well-established problem in the domain of software learning, particularly for complex, feature-rich software \cite{mcgrenere2000, mackamul2023}. Users can face significant difficulties in discovering and exploring further features and practices beyond those they are used to applying \cite{mcgrenere2000, cockburn2014, mackamul2023}. We argue that there are parallels and lessons which could be drawn from the current literature to help understand how this issue manifests in the context of GenAI tools. 

One explanation for what we observed could be that our participants' experiences uniquely reflect the issues faced by users transitioning from traditional, graphic user interface and feature-based software applications, to tools with generative and far more powerful capabilities. A software application is traditionally developed with pre-defined purposes, functionalities, and workflows by the developers \cite{goransson1986, nouwens2018}, which may or may not fully meet the needs of the users. Knowledge workers typically engage in high levels of application-switching in order to address the needs of a particular task, particularly if applications do not provide the functionalities they need to access \cite{jahanlou2023}. Challenges with discovering novel GenAI use may therefore be an issue of users navigating with these previous constraints, and not recognising the potential for GenAI's capabilities to be extended beyond the existing modalities, tasks, and domains that they currently use it for. For example, work by Renom \cite{renom2023} highlights that users' abilities to re-purpose and craft alternative practices within a tool can often be limited by their biases and their own prior knowledge of other digital environments, as well as interface cues. It may therefore have been challenging for our participants to learn to recognise that GenAI tools do not share the constraints common to other software. In such cases, improving education and understanding of the underlying technological model lowering barriers to exploration could help to mitigate these potential fallacies in understanding. 

\subsubsection{Challenges to continuous learning around GenAI}
\update{Beyond discoverability, exploring GenAI's capabilities can be challenging because it requires continuous, long-term effort. The possible applications of GenAI tools are both extensive and rapidly evolving. Users must therefore contend with updates to existing models can lead to differences and inconsistencies in outputs and task suitability, as well as make sense of multiple competing GenAI tools and increasingly specialised applications of such tools \cite{kotturi2024, bansal2019}. What's more, in workplace contexts, where there are often time pressures and competing work demands, it can also be difficult for users to maintain learning on top of their existing workloads \cite{lemmetty2020a}.} 

\update{Knowledge sharing between peers and colleagues is therefore especially important as a learning process that enables users to navigate resource constraints \cite{lemmetty2020a, tynjala2008}. In line with theories of communities of practice (CoP) \cite{wenger2000}, learning can be understood as a socially situated process in which groups of individuals, united by a shared domain, develop and refine practices through sustained participation and interaction. Consistent with prior work \cite{liu2026}, our findings show that much of the learning, exploration, and understanding of GenAI practices is socially negotiated. Because there are so many possibilities for interacting with, and adapting GenAI tools to existing or novel workflows, users benefit from the experiences shared by others to help them make sense of these possibilities. However, our findings also extend CoP theories. While CoP emphasises active participation as central to shaping and reinforcing individuals’ membership within a community, we find that, users proactively conceal traces of GenAI usage as a way to reinforce their identities as domain experts. In other words, rather than sustaining interactions to reinforce membership and identity as experienced GenAI users, for some users, competing identity priorities and the desire to emphasise own `domain expertise' drive behaviours which, long-term, threaten the sustainability of GenAI-focused communities and discourse around GenAI use (see Section \ref{subsec:disc_risks_non_disclosure}}). Indeed, our participants emphasised that much of their awareness of GenAI tools is predominantly driven current popular discourse and heightened media representation, and that such interest is not necessarily maintained once usage becomes normalised \cite{widder2024}. It is therefore important to consider the resources and structures which organisations could implement to support AI literacy not only as a one-time phenomenon, but a continuous process of exploration, maintenance, and transfer for both individuals and the communities they reside in \cite{feng2026}.

\subsection{Implications}
\subsubsection{Organisations}
For organisational leaders and managers, we emphasise that the development of an open, social workplace environment is one of the most important conditions for supporting effective GenAI learning. \update{To achieve this, managers should establish effective channels for feedback around GenAI usage, and be open-minded to receiving diverse perspectives from employees \cite{morrison2000}. In particular, management should take care to recognise how fear of external judgement dictates behaviours such as GenAI hiding, and be cautious in navigating the power dynamics between employees and managers which can exacerbate these issues \cite{morrison2000}. Engaging in discourse around GenAI use in informal contexts, and reducing the distance between managers and employees, can also provide individuals with greater psychological safety that can better encourage social learning and improve awareness across the hierarchy of the organisation \cite{edmondson1999}}.

We also highlight that passive or implicit allowance of GenAI use may not necessarily encourage discourse. As discussed in Section \ref{subsubsec:negative_impact_KS}, despite observations of GenAI use, individuals may still refrain from engaging in knowledge sharing out of concern of potential negative judgement and repercussions from others \cite{zhang2025, liu2026}. \update{In contexts where there is uncertainty, employees can often opt for risk-averse strategies regarding disclosure which exacerbate `organisational silences' and willingness to report on concerns or problems \cite{morrison2000}.} This could be of particular concern when onboarding new employees, as a lack of openness and discussion about the use of GenAI can make it challenging for learning and experience to be circulated. Additionally, in blended work environments, AI acts not only as a tool but as a co-worker and co-author, which amplifies the stakes of disclosure, credibility, and ownership \cite{constantinides2025}. 

\update{To address these issues, we suggest that organisations should be explicit and targeted in cultivating communities of practice around GenAI use as a priority \cite{wenger-trayner2023}. Considering the volume and rapidity of GenAI developments, and the scale of possible GenAI use cases, reward incentives could be established to support the different functions needed to support such continuous learning. For example, organisations could address the issue of maintaining awareness of new developments by providing funding for certain employees to trial GenAI tools, experiment with workflows, and develop specialised solutions. Organisations could also draw explicit attention to the issue of GenAI disclosure, and implement tools and templates to support disclosure from other contexts, such as academia or journalism (e.g. \cite{ahmetoglu2026, parast2025, weaver2024a, bogucka2025}), and tailor them to suit the specific contexts where disclosure is mandated within the organisation or when working with external stakeholders. Beyond simply implementing tools for disclosure, encouraging users to reflect on the appropriateness and design of disclosure tools, can also provide an important opportunity for managers to gain a deeper understanding of how GenAI tools are used by workers, how they believe such information should be communicated, and to collectively negotiate and develop a shared understanding of standards for GenAI disclosure within their community.}

\subsubsection{Design}
\update{Echoing prior work on software learning \cite{kiani2020, lemmetty2020a, liu2026, xia2025}, our study finds that GenAI users benefit from shared experiences and artefacts produced by others. Such knowledge helps users uncover new possibilities for GenAI use and adapt their own practices. However, the key challenge which designers must overcome is to identify the level of abstraction and disclosure users are comfortable with sharing regarding their GenAI usage, under what contexts, and with whom. To address this challenge, we suggest that designers should consider both existing interventions to support the communication of GenAI usage patterns, as well as the implementation of socio-technical systems which can encourage repeated engagement in sharing behaviours.} 

\update{For example, our participants commonly reflected on the idea that sharing of GenAI-related interaction knowledge can be challenging without first-hand experience, due to the unpredictable nature of GenAI outputs. Existing documentation practices of GenAI users on social media forums reflect this complexity, with users often sharing prompts, partial transcripts, screenshots, or highly detailed descriptions of processes outlining workflows, GenAI tools used and corresponding settings, which can be challenging to parse and interpret in context \cite{liu2026, tang2026}. We suggest that, initially, designers should focus on readily available traces of GenAI interactions, such as prompts, which could be more readily shared or collaboratively developed between users \cite{liu2026, feng2024}. However, given the evolving nature of GenAI tools \cite{bansal2019}, the framing around prompt-sharing should be better communicated, not necessarily for the purpose of replicating outcomes through reuse, but instead in revealing the metacognitive strategies needed to structure and articulate tasks and the interactions needed to reach a desired outcome \cite{tankelevitch2024, dalsgaard2025}. To address the socio-technical challenges of stigma around GenAI usage, interventions which foreground positive social incentives may help mitigate concerns around sharing. For example, appreciation systems, widely used in open-source communities and platforms such as Wikipedia \cite{spiro2021}, allow users to give positive feedback and visible signals of appreciation in response to contributions. Recent systems such as HugReports \cite{khadpe2025} have demonstrated that, embedding such appreciation systems within a user's development workflow and allowing them to signal appreciation for individual packages and modules created by other developers can help contributors feel recognised for their work, and also to serve as feedback for what is useful and what is not. Extending such approaches to shared GenAI artefacts, such as individuals contributing useful prompts, tool integrations, and workflows may both support knowledge exchange and increase the visibility of social rewards associated with sharing.} 

\update{We highlight, however, that much of the existing interactions with GenAI use remain private and individual \cite{xiao2026}, and thus more research is needed to identify where there could be useful overlaps and opportunities for collaborative GenAI usage that could support these forms of interactions and system implementations. Designers could follow the approach adopted by Giannisakis et al. \cite{giannisakis2022} to develop a better understanding of the levels of abstraction for sharing and disclosure which users are comfortable with, and what design features are needed to support this method of sharing. This topic remains an open space for exploration which will likely continue to evolve as new patterns for GenAI usage and opportunities for user collaboration emerge.}

\subsection{Limitations and Future Work}
\edit{While our work provides important insights into how current adopters of GenAI negotiate with the concept of GenAI literacy in the workplace, there still remain several open questions for future works to explore. First, our work identifies the important role of other users in supporting discoverability of GenAI functionalities and applications to work, but it is not known} to what extent which GenAI \note{technologies could be applied to fulfill or replace this role, and what the consequences might be.} Many current commercial GenAI tools do not yet incorporate clear, explainable AI principles to comprehensively support user understanding of how they may be used in context. Though explainable AI principles can be embedded into design \note{to support the goal of learning, we suggest that the social influence of others on the practice of GenAI usage is likely to persist regardless. Our findings suggest that, regardless of whether or not people actually learn from their colleagues, the act of sharing one's software knowledge can still be an important act for accumulating social credit and signalling domain expertise,} and people's understanding of which GenAI competencies matter (if they matter at all), is still influenced by others' attitudes. \edit{Greater long-term research on the implications of applying GenAI tools to support GenAI literacy, and the extent to which such observations may manifest in future generations, will be needed to better support our understanding of how we should engage in technological learning.}

A further area for exploration is to consider the influence of individual variables and demographics on how users engage in learning. Though we aimed to capture a diverse sample of participants in our study, we did not explicitly examine the influence of individual factors which could also play a role in how susceptible an individual is to specific external or social influences. For example, prior literature suggest that women possess lower self-efficacy, i.e., confidence in their own technical knowledge \cite{ying2021}, and gender differences in motivation and types of knowledge hiding that is also impacted by the perceived social roles of both genders \cite{andreeva2023}. Future work could consider exploring potential associations between factors such as gender, cultural background, technical expertise, or personal variables such as self-efficacy, and personality variables, as well as organisational culture, and intention to share knowledge. 

Finally, the findings of this study is bounded by the characteristics of our sample – namely, participants who were already willing adopters of GenAI technology, including those who may be using GenAI against organisational recommendations. \update{As such, common concerns around GenAI use, including issues of fairness and attribution, which are more prevalent concerns for those reluctant to use GenAI \cite{rana2024}, were not identified as strongly as issues of work legitimacy, expertise, and reputation. While our participants reflected on who should use GenAI under different contexts, there was less emphasis on the evaluation and decision-making process surrounding when GenAI should be used, a theme which is often more salient in individuals who engage in GenAI non-use \cite{cha2025}, or in contexts where GenAI adoption is enforced \cite{stewart2025}. While these perspectives fall outside the scope of this study, incorporating them in future works is important for developing a more comprehensive account of GenAI literacy and user–technology relationships.}

\section{Conclusion}
This paper extends current discussions on GenAI literacy through an empirical exploration into how knowledge workers outside of formal education define key GenAI competencies and strategies for learning. We found that users value broad awareness of GenAI's capabilities due to the complexities of the constantly evolving field and to enhance their own competitiveness. Furthermore, learning from the experiences of others is crucial for maintaining this awareness. However, concealing `obvious' \edit{cues} of GenAI usage is also valued as a competency, not only to avoid stigma, but as a way of re-affirming users' own domain expertise. We contribute to existing knowledge on GenAI hiding behaviours by highlighting \edit{why users may be motivated to engage in hiding}, and that these actions can promote a lack of transparency and critical reflection on GenAI use. To mitigate hiding behaviours, we suggest that designers enhance the visibility of user-generated knowledge \note{when using} GenAI tools, and that organisations can better emphasise the benefits of maintaining collaborative learning and open discussions to support responsible, transparent use. 

\section{AI Disclosure Statement}
The authors acknowledge the use of Generative AI tools (ChatGPT 5.2) to support language improvements across the manuscript (to refine phrasing and improve readability), and to support brainstorming and ideation based on author's instructions. All outputs were reviewed and verified by the authors. All substantive ideas, descriptions, summaries, and arguments were developed by the authors, who take full responsibility for the final content. The use of generative AI followed relevant ethical guidance. This disclosure statement was created using DAISY (available at: https://dryoanaahmetoglu.github.io/daisy-disclosure/).